\documentclass[review,authoryear,12pt]{elsarticle}
\usepackage[hmargin=1in]{geometry}

\usepackage{amssymb}
\usepackage{amsmath}

\usepackage[hyphens]{url}
\usepackage{color}

\journal{Icarus}

\begin{document}
\begin{frontmatter}

\title{Simulations of Titan's Paleoclimate}
\author{\textit{Published in Icarus 243 (2014) 264--273}\\ \vspace{10pt}
Juan M. Lora$^1$, Jonathan I. Lunine$^2$,
Joellen L. Russell$^1$, Alexander G. Hayes$^2$}
\address{$^1$Department of Planetary Sciences, University of Arizona, Tucson, AZ 85721\\
$^2$Department of Astronomy, Cornell University, Ithaca, NY 14853}


\begin{abstract}
We investigate the effects of varying Saturn's orbit on the atmospheric circulation and surface methane distribution of Titan. Using a new general circulation model of Titan's atmosphere, we simulate its climate under four characteristic configurations of orbital parameters that correspond to snapshots over the past 42 kyr, capturing the amplitude range of long-period cyclic variations in eccentricity and longitude of perihelion. The model, which covers pressures from the surface to 0.5 mbar, reproduces the present-day temperature profile and tropospheric superrotation. In all four simulations, the atmosphere efficiently transports methane poleward, drying out the low- and mid-latitudes, indicating that these regions have been desert-like for at least tens of thousands of years. Though circulation patterns are not significantly different, the amount of surface methane that builds up over either pole strongly depends on the insolation distribution; in the present-day, methane builds up preferentially in the north, in agreement with observations, where summer is milder but longer. The same is true, to a lesser extent, for the configuration 14 kyr ago, while the south pole gains more methane in the case for 28 kyr ago, and the system is almost symmetric 42 kyr ago. This confirms the hypothesis that orbital forcing influences the distribution of surface liquids, and that the current observed asymmetry could have been partially or fully reversed in the past. The evolution of the orbital forcing implies that the surface reservoir is transported on timescales of $\sim$30 kyr, in which case the asymmetry reverses with a period of $\sim$125 kyr. Otherwise, the orbital forcing does not produce a net asymmetry over longer timescales, and is not a likely mechanism for generating the observed dichotomy.
\end{abstract}

\end{frontmatter}

\section{Introduction}
The drastic hemispheric asymmetry observed in the distribution of Titan's lakes has been suggested to be the consequence of asymmetric seasonal forcing. \citet{Aharonson09} hypothesized that net drying of the pole subject to shorter but more intense summers would lead to this asymmetry on a timescale comparable to the period of Saturn's orbital variations, $10^4$ years. This would imply that $\sim$30 kyr in the past, when the orbital parameters produced a longitude of perihelion passage near the northern summer solstice, the asymmetry would have been reversed.
\par
General circulation models (GCMs) of present-day Titan have shown that the atmosphere effectively transports methane poleward, drying the equatorial regions \citep{Rannou06,Mitchell08} where dune fields are located \citep{Radebaugh08}. \citet{Schneider12} used a GCM that included diffusion of surface liquids to further show that increased precipitation at the north pole, which undergoes milder but longer summers in the current epoch, results in buildup of its surface reservoir, creating such an asymmetry.
\par
Recent clouds \citep{Schaller09} and rainstorms \citep{Turtle11b} have been observed at Titan's low latitudes, strengthening the possibility that equatorial fluvial surface morphologies, like the channels observed by Huygens \citep{Tomasko05} and washes seen by Cassini RADAR \citep{Lorenz08}, are the result of seasonal rainfall as observed in the present, despite the prevalence of desert-like conditions near the equator.
\par
In this paper, we investigate the effects that changes in Saturn's orbit have had on Titan's climate in its recent geologic history. With a general circulation model, we simulate Titan's climate during four characteristic orbital configurations that occurred in the past 42 kyr. In addition to the present day, the times chosen correspond to the maximum (14 kyr ago) and minimum (42 kyr ago) values in the recent variation of Saturn's orbital eccentricity, and a midpoint between them (28 kyr ago). The timespan of these snapshots also corresponds to a large variation of longitude of perihelion (see Table \ref{table:parameters}). All of these orbital elements are representative of those over the past Myr, as can be seen in Fig.~\ref{Fig:orbit_params}; a description of their computation is provided in the Appendix.
\par
Of particular interest is the distribution of surface methane over the past 40 millennia: Where have liquids accumulated over this time? Are there marked differences in the distribution of surface temperatures? Has the circulation of the atmosphere changed significantly?

\begin{table}
\centering
\footnotesize
\caption{Simulation orbital parameters}\label{table:parameters}
\begin{tabular}{ccccc}
\hline
Time b.p. (kyr) & Eccentricity & $L_S$ of perihelion (deg)& Obliquity (deg)& Semi-major axis (AU)\\
\hline
0 & 0.054 & 277.7 & 26.7 & 9.54 \\
14 & 0.081 & 202.6 & 26.8 & 9.53 \\ 
28 & 0.054 & 121.4 & 28.2 & 9.55 \\
42 & 0.014 & 330.6 & 27.7 & 9.58 \\ 
\hline
\end{tabular}
\
\end{table}

\begin{figure}[h!]
	\begin{center}
	\includegraphics[width=0.5\textwidth]{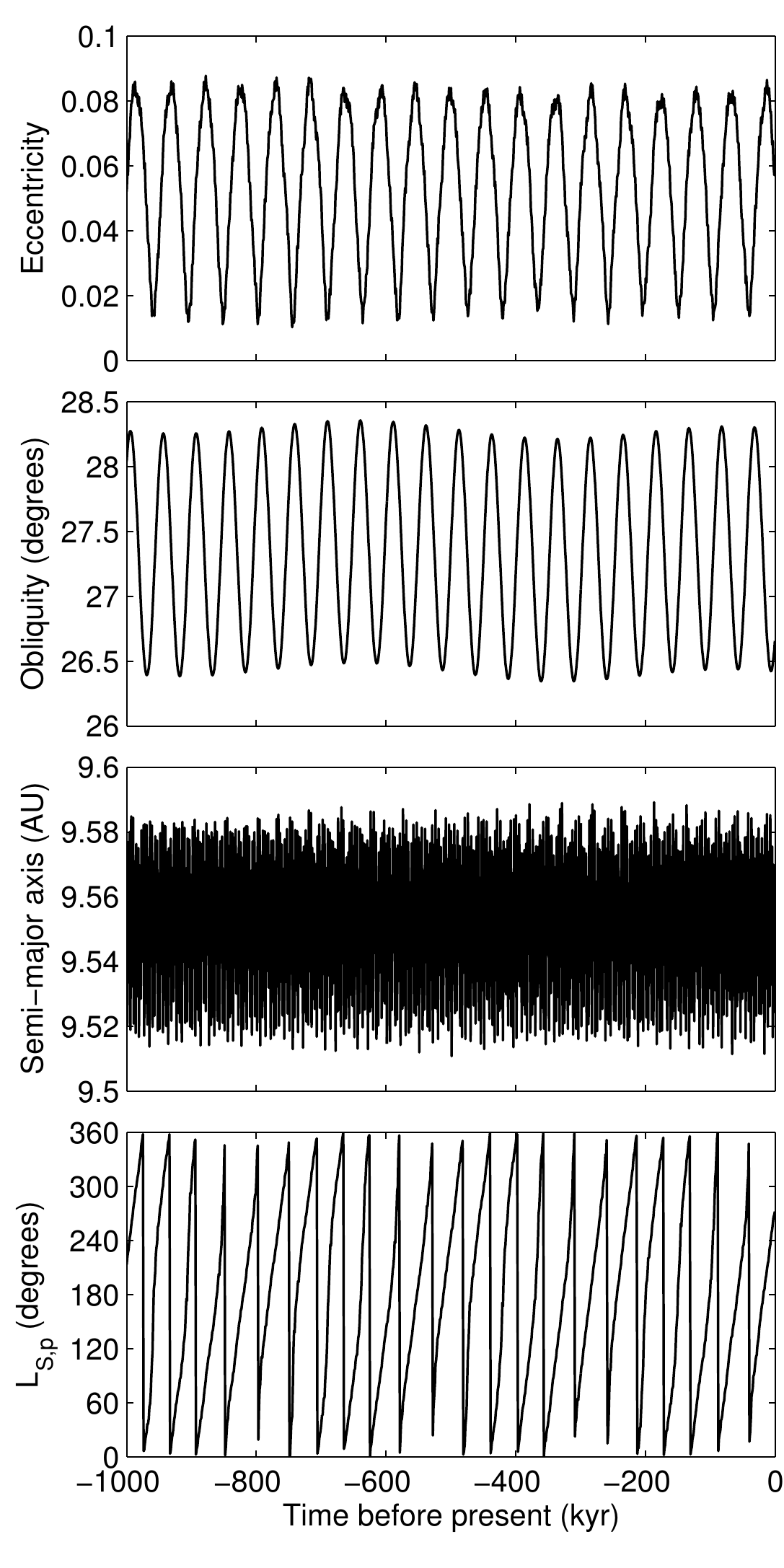}
	\caption[Orbital parameters]{Variation of Titan's effective orbital parameters over the past million years. From top to bottom: orbital eccentricity, obliquity, semi-major axis, and solar longitude of perihelion. See Appendix for details.\label{Fig:orbit_params}}
	\end{center}
\end{figure}

\section{Model description}
\subsection{Dynamical core}
The model integrates the hydrostatic primitive equations, in vorticity-divergence form, using the Geophysical Fluid Dynamics Laboratory's (GFDL's) spectral transform dynamical core \citep{Gordon82}. The core is run with triangular truncation, leapfrog time integration and Robert filter, and eighth-order hyperdiffusion.  In this study, the resolution is T21 (approximately equivalent to 5.6$^\circ$ horizontal resolution) to minimize computational time. 
\par
The vertical coordinate, here using 25 levels, is a hybrid that smoothly transitions from unequally-spaced sigma coordinates near the planetary surface to pressure coordinates near the model top, at approximately 0.5 mbar. This choice allows for the inclusion of topography, though this is not presently used.  It also provides higher resolution in the boundary layer than would an evenly-spaced sigma coordinate. 
\par
Energy and moisture are corrected by a multiplicative factor to guarantee conservation by the dynamics. An additional multiplicative factor correction is made to the surface liquid reservoir to ensure that total methane is conserved in the model after exchange between surface and atmosphere, during which small numerical errors otherwise produce a systematic sink for methane. The details of the exchange processes and the reservoir are described further below.

\subsection{Radiation}
Because atmospheric dynamics are in essence driven by radiative heating, and in particular because Titan's seasonal weather is controlled by fluxes from the surface, which in turn respond to insolation, the accuracy of radiative transfer is of primary importance.  Therefore, we developed a nongray, multiple scattering radiative transfer module that takes advantage of relevant data from the Cassini-Huygens mission. Titan's diurnal and seasonal cycles are fully accounted for in the computation of the top-of-atmosphere insolation.
\par
The radiative transfer uses delta-Eddington and hemispheric-mean two stream approximations \citep{Toon89} for computation of solar (visible and near-infrared) and thermal infrared fluxes, respectively. Radiative fluxes between atmospheric layers are computed using layer scaled optical properties---extinction optical depth, single scattering albedo, and asymmetry parameter---combined from individual sources of opacity. These sources (including multiple scattering in the solar spectrum) are methane, haze, and Rayleigh scattering for solar radiation, and collision-induced absorption (CIA) from combinations of N$_2$, CH$_4$ and H$_2$ pairs, as well as molecular absorption by CH$_4$, C$_2$H$_2$, C$_2$H$_4$, C$_2$H$_6$, HCN, and absorption by haze, in the infrared. The profiles of the stratospheric molecular species are fixed to realistic values \citep{Vinatier07}.
\par
A combination of exponential-sum fits and correlated $k$ coefficients \citep{Lacis91,Fu92} is used to define gas broadband transmittances for each layer. In the solar portion of the spectrum, opacities for methane at wavelengths short of 1.6 $\mu m$ are calculated with fits to transmittance from DISR absorption coefficients \citep{Tomasko08a}, while methane opacities between 1.6 and 4.5 $\mu m$ are accounted for using correlated $k$ coefficients calculated from HITRAN line intensities \citep{Rothman09}. In the thermal infrared, CIA opacities are calculated with fits to transmittance from HITRAN binary absorption cross-sections \citep{Richard11}, while molecular absorption is treated with correlated $k$ coefficients from temperature- and pressure-corrected \citep{Rothman96} HITRAN line intensities. Both the fit parameters and coefficients (with associated weights) are computed offline for the range of temperatures and pressures relevant to Titan. 
\par
The effects of haze are incorporated using the optical parameters from DISR at solar wavelengths, and extinction coefficients determined from Cassini/CIRS data between 20-560 cm$^{-1}$ \citep{Anderson11} and 610-1500 cm$^{-1}$ \citep{Vinatier12} at infrared wavelengths. Multiple scattering is accounted for using the haze single scattering albedo and asymmetry factors, computed from phase functions, at solar wavelengths \citep{Tomasko08b}.  The haze is assumed to be a pure absorber in the thermal infrared \citep{Tomasko08c}. The haze is also not dynamically coupled, and is thus assumed to be horizontally homogeneous. Haze coupling has been shown to amplify wind speeds and temperature contrasts in the stratosphere \citep{Rannou04}, but this effect should be of little consequence for the lower troposphere, as the most significant haze accumulation occurs over winter high latitudes in polar night, and as such the insolation distribution is not affected.

\subsection{Moist processes}
Parameterization of large-scale condensation (LSC), which occurs when the relative humidity in a grid box reaches or exceeds 100\%, is included. Resulting methane condensate is allowed to re-evaporate as it falls through the atmosphere below, such that rain only reaches the surface when the atmosphere below the condensing grid box becomes saturated. 
\par
The saturation vapor pressure is calculated either over methane-nitrogen liquid \citep{Thompson92} or methane ice \citep{Moses92}. In the former case, we assume a fixed latent heat of vaporization (with a reference saturation vapor pressure of 106.0 mbar at 90.7 K) and a constant nitrogen mole fraction of 0.20. The transition between liquid and ice is chosen to occur where the vapor pressure curves intersect, at approximately 87 K. This represents a mild suppression of the freezing point below the triple point of pure methane due to the binary liquid. Where condensation occurs, only the latent heat of vaporization is used, ignoring the $\sim$10$\%$ difference with that of sublimation; this is equivalent to assuming that only rain is produced, which avoids the need to also include the liquid-ice transition for energy balance. Given the uncertainties in the composition of the condensate and the expectation that surface precipitation is liquid \citep{Tokano06}, this assumption does not add to the uncertainty in the simulations.
\par
The effects of ethane are not modeled. Because of its much lower vapor pressure, ethane would have only an indirect effect in reducing the methane evaporation rate by its fractional abundance in the liquid. Unless ethane dominates strongly over methane, this is not a big effect, and would also impact all simulations equally. Furthermore, though ethane may be incorporated in condensates in the troposphere, its effect should be small and previous studies of the humidity profile have also neglected it \citep[i.e.][]{Tokano06}.
\par
In addition to LSC, a quasi-equilibrium convection scheme is included, as in previous studies of Titan's methane cycle \citep{Mitchell06,Schneider12}. Moist convection is parameterized with a simplified Betts-Miller scheme \citep{Frierson07,O'Gorman08}, where a convectively unstable column of atmosphere is relaxed toward a moist pseudoadiabat, wringing out excess liquid.  This approach assumes that droplet nucleation is always possible, and that droplets quickly become large enough to fall to the ground; in this scheme, all resulting rain reaches the surface.

\subsection{Planetary boundary layer}
Vertical diffusion due to boundary layer turbulence is parameterized with a K-profile scheme. Heat, moisture, and momentum diffusivities are calculated as a function of both height within the boundary layer and stability functions determined from Monin-Obukhov similarity theory, as well as surface friction velocity. The height of the boundary layer is set at a critical Richardson number (here set to 1.0) for stable and neutral conditions near the surface, or at the level of neutral buoyancy for surface parcels in the case of unstable conditions, and is thus variable depending on the environment. 

\subsection{Surface model and surface fluxes}
The atmospheric component of the GCM is coupled to a land model that calculates temperatures in the surface and sub-surface. Fluxes of sensible and latent heat, as well as radiative fluxes (including emitted thermal flux), affect the top-most surface layer. Conduction transports heat between soil layers. We use 15 layers of variable thickness, and assume a constant thermal conductivity and volumetric heat capacity of 0.1 Wm$^{-1}$K$^{-1}$ and 1.12$\times$10$^6$ Jm$^{-3}$K$^{-1}$, respectively, values appropriate for a ``porous icy regolith" \citep{Tokano05}.
\par
The surface model also tracks the presence of surface liquid with a ``bucket'' model: Liquid from precipitation is allowed to accumulate, and is then available to evaporate. Below a threshold amount of liquid, a linearly decreasing availability factor roughly parameterizes infiltration and sub-grid scale ponding, limiting evaporation. Infiltration is not explicitly modeled, and the effects of surface liquid on soil properties are not currently included. The total amount of methane in the system (surface liquid and atmospheric moisture) is conserved. 
\par 
Surface fluxes of momentum, heat, and vapor are computed using bulk aerodynamic formulae (with the above-mentioned modification to vapor flux), with drag coefficients from Monin-Obukhov similarity theory. The roughness length and gustiness parameters in this module are set to 0.5 cm and 0.1 m s$^{-1}$, respectively \citep{Friedson09,Schneider12}.

\section{Results}
\subsection{Insolation distributions}
\begin{figure}[h!]
	\begin{center}
	\includegraphics[width=0.8\textwidth]{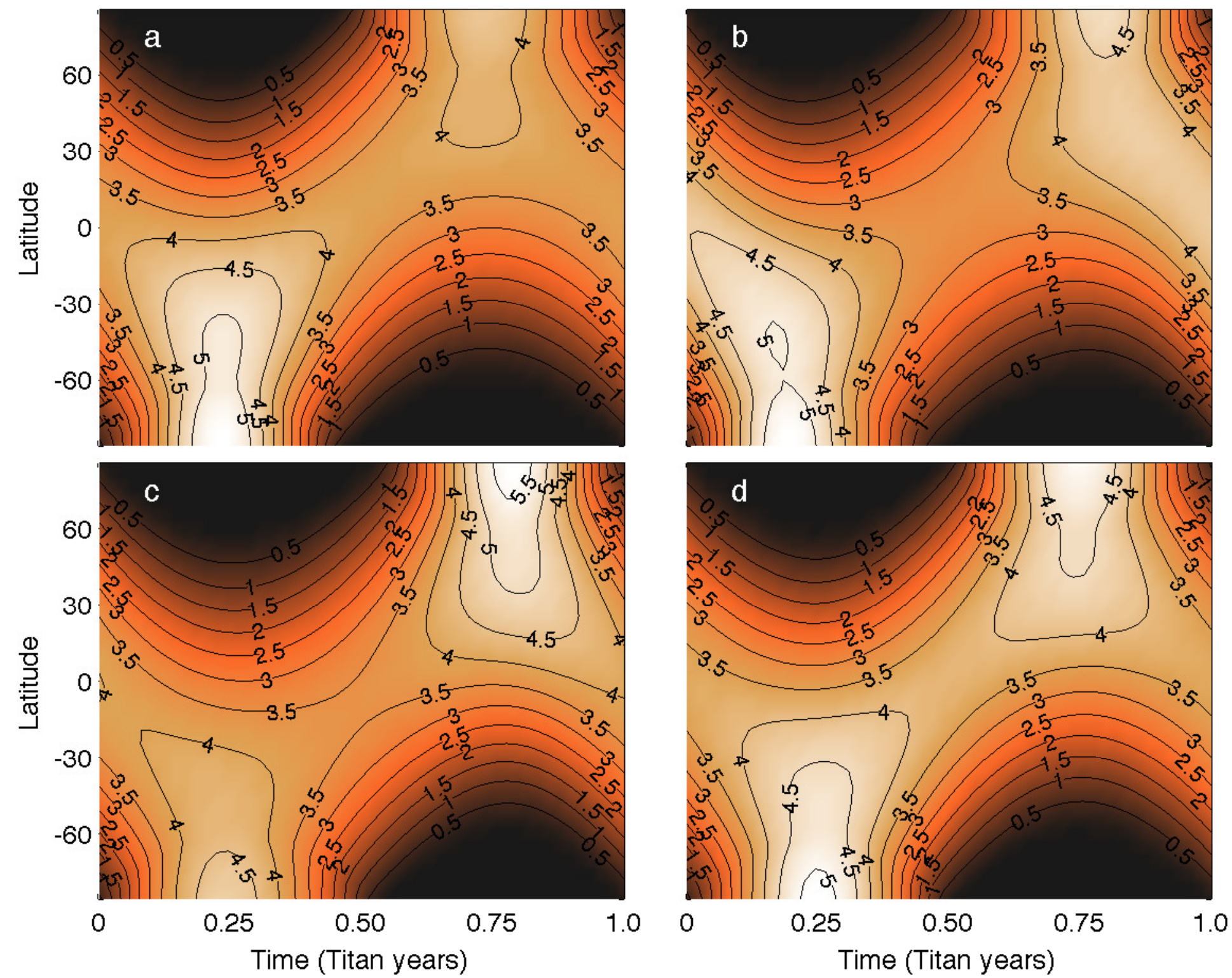}
	\caption[Insolation Distributions]{Insolation distributions (W/m$^2$) at the top of the model atmospheres for the four cases considered: (a) Present-day insolation distribution; (b) distribution for orbital parameters corresponding to 14 kyr ago; (c) distribution corresponding to 28 kyr ago; (d) distribution corresponding to 42 kyr ago. \label{Fig:sw_toa}}
	\end{center}
\end{figure}

The direct consequence of varying the orbital elements is a variation in the distribution of solar flux incident at the top of Titan's atmosphere. This distribution is responsible for seasonal effects as well as any differences that might arise from the different orbital configurations. It is worth first comparing the resulting distributions (Fig. \ref{Fig:sw_toa}) independently of other results from the GCM.

\par
The present-day scenario produces a familiar distribution of diurnal-mean solar flux \citep{Aharonson09}, where the southern summer is shorter but more intense than the northern summer, and therefore the southern polar night is longer than its northern counterpart. This is due to the occurrence of southern summer coinciding quite closely with Saturn's perihelion, in the planet's currently moderately eccentric orbit.
\par
The distribution with orbital elements from 14 kyr ago displays a pronounced difference in the insolation received at the equator between the opposite equinoxes, owing to perihelion occurring closer to the southern vernal equinox, and to the higher orbital eccentricity. While the southern pole still receives slightly more insolation during summertime, the enhancement versus the northern summer pole is $\sim$10$\%$, less than half that of present day. The mid-latitudes also receive summertime insolation that is relatively enhanced versus the poles in comparison with the present-day case.
\par
The eccentricity 28 kyr ago is the same as at present, but in this scenario perihelion occurs close to northern summer. In addition, a slightly higher obliquity causes a further enhancement of summertime insolation of the poles versus other latitudes. At its maximum the north receives about 20\% more insolation than the south. This scenario is in some ways the latitudinal opposite to present-day, though the longitude of perihelion is not exactly 180$^{\circ}$ away from its present value \citep[that happened $\sim$31.5 kyr ago;][]{Aharonson09}. There is a slight enhancement of insolation during northern spring versus southern spring, as compared to present-day. 
\par
Finally, the distribution from 42 kyr ago is the most latitudinally symmetric, because the eccentricity at this time reaches a minimum, and perihelion again occurs close to equinox. Thus, south and north polar regions have seasons of similar intensity and duration, with only a minimal difference between spring and autumn insolation at low latitudes.

\subsection{Model spin-up}
The model is first spun up from rest, including the present-day seasonal cycle, and imposing a deep surface reservoir of methane, allowing the atmosphere to moisten until the humidity reaches a steady state. This occurs after about ten Titan years of integration ($\sim$290 years); we continue the spin-up simulation for an additional ten Titan years. Then, we replace the deep surface reservoir with one that is three meters of methane globally and allowed to dry up, and this setup is allowed to evolve under the different orbital conditions. Some dependence on the initialization is expected and apparent in the early years of the simulations, but disappears with sufficient simulation time (see for example Fig.~\ref{Fig:sum_surf_liq}).

\subsection{Present-day atmosphere}
We briefly describe a few aspects of the simulated present-day atmosphere to both show the validity of the model results and provide a basis for comparison of the results from the other experiments.

\begin{figure}[h!]
	\begin{center}
	\includegraphics[width=0.8\textwidth]{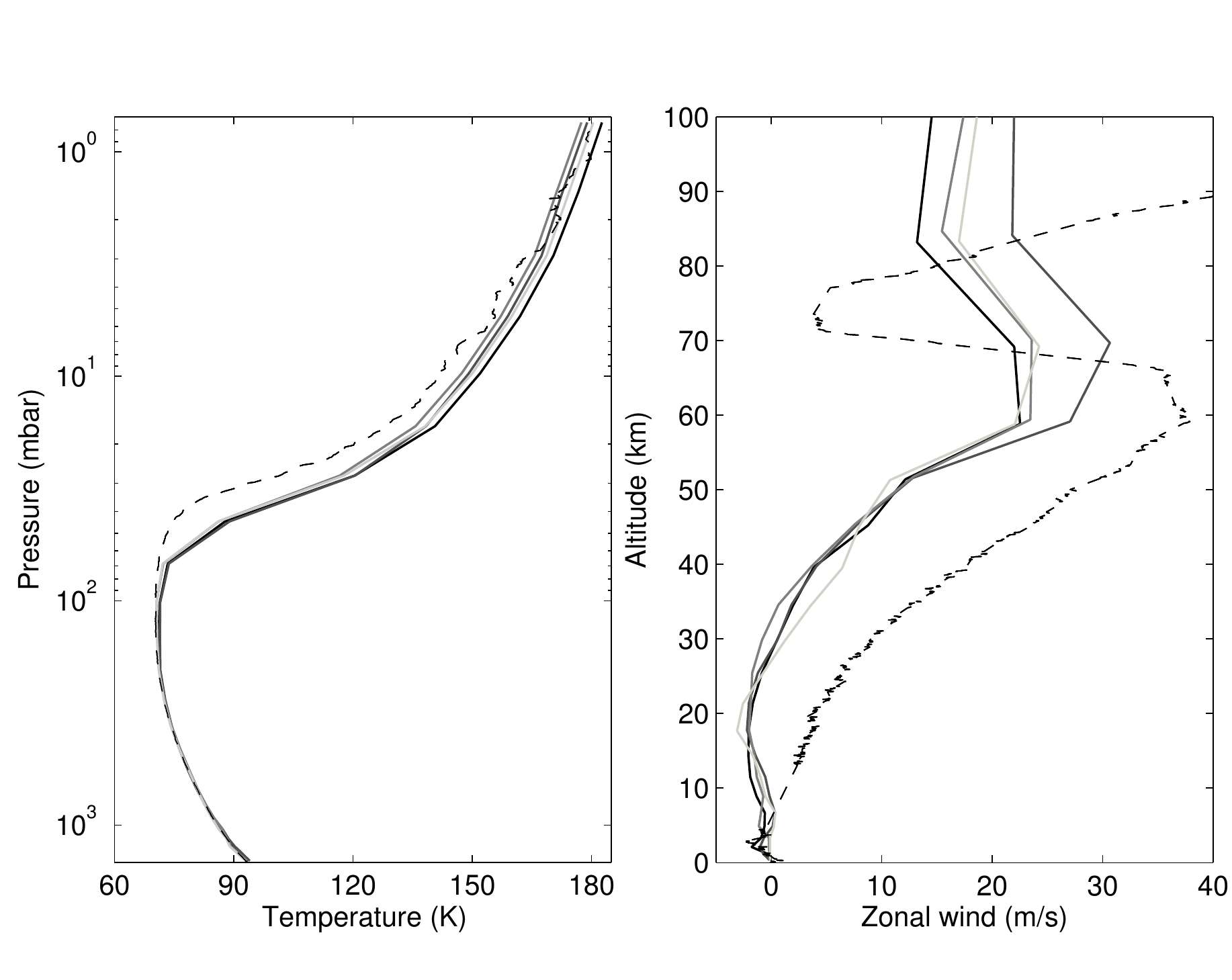}
	\caption[Temperature/Wind]{Left: Modeled atmospheric temperatures (solid lines) at the season and latitude of the Huygens probe's descent, compared to the observations (dashed line) from HASI \citep{Fulchignoni05}. Right: Modeled zonal wind versus Huygens' observations \citep{Bird05}. Solid lines correspond, from darkest to lightest, to present-day, 14 kyr ago, 28 kyr ago, and 42 kyr ago simulations, respectively. Note the different vertical scales between the panels. \label{Fig:temp_wind}}
	\end{center}
\end{figure}

\par
Firstly, the atmospheric temperature profile, at equatorial latitudes, is in generally good agreement with the profile measured during Huygens' descent (Fig. \ref{Fig:temp_wind}). The stratosphere occurs at slightly higher pressures than observed---alternatively, stratospheric temperatures are too warm---possibly due to the lower vertical resolution in this region and the low model top, though this is an effect also encountered in other full-radiation Titan models \citep{Friedson09,Lebonnois12} and may be due to uncertainties in CIA or methane absorption optical properties. 
\par
At high latitudes, the tropopause temperature can reach as low as 67.5 K (see Supplementary material, Fig.~S1), and the temperatures oscillate seasonally by $\sim$3 K. In the stratosphere, horizontal temperature gradients are only a few degrees, in contrast to observations \citep{Achterberg11}. Both the low model top and the exclusion of haze transport probably contribute to this excess of horizontal temperature homogeneity.
\par
Modeled zonal winds in the troposphere agree reasonably with observations \citep{Bird05}. Low-altitude zonal winds tend toward westward motion, but the integrated troposphere is superrotating. The stratosphere also superrotates, though the profile does not agree with observations. However, the low model top and lack of haze transport are expected to strongly affect the top several model layers, so significantly weaker-than-observed winds should be expected. The natural development of superrotation in the model, though significantly lower in magnitude than observed in the stratosphere \citep{Achterberg08}, suggests robust model dynamics.

\subsection{Precipitation and surface reservoirs}\label{Sec:precip}
In all of our simulations, summertime polar rain is a robust and consistent feature. Though Titan's slow rotation period of about 16 days results in its mean meridional circulation extending into high latitudes, the region of upwelling near the surface associated with the Hadley cell is limited in its poleward extent (Supplementary material, Fig.~S1). (This has been shown to be due at least in part to latent heat release \citep{Mitchell06}.) Polar precipitation is thus associated with a small region of upwelling at the pole that develops concurrently with the Hadley upwelling arriving at its highest latitude ($\sim$50$^{\circ}$). In the present-day case, northern summer precipitation tends to begin approximately three (Earth) years before solstice, though it ramps up significantly one year before solstice (Supplementary material, Fig.~S5).

\begin{figure}[h!]
	\begin{center}
	\includegraphics[width=0.8\textwidth]{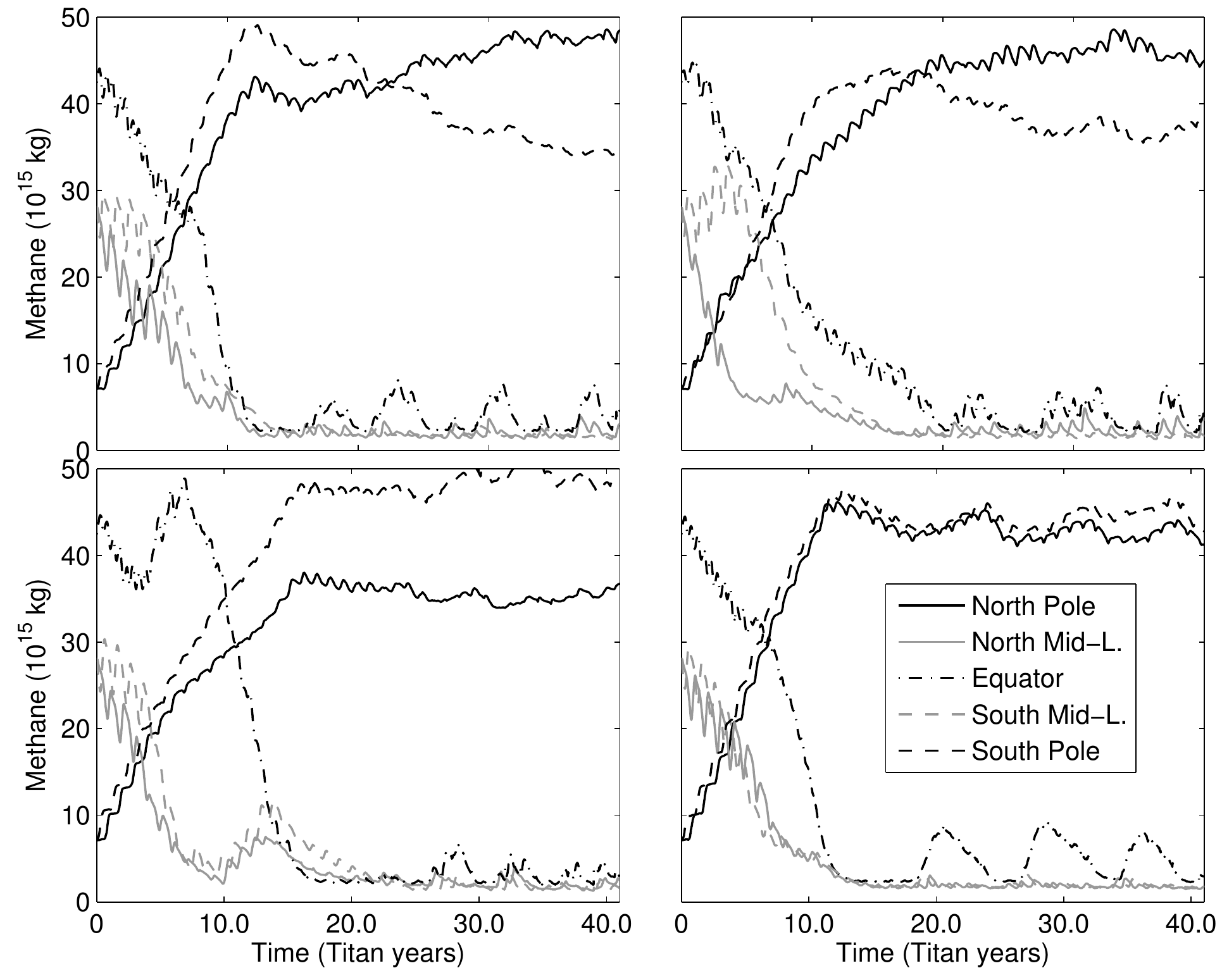}
	\caption[Surface Liquids]{Total liquid methane reservoir at the surface in five latitude bins corresponding to 90N--60N, 60N--30N, 30N-30S, 30S-60S, and 60S-90S, for the four simulations. Top: Present-day (left) and 14 kyr ago (right). Bottom: 28 kyr ago (left) and 42 kyr ago (right). The legend for all plots is shown in lower right.\label{Fig:sum_surf_liq}}
	\end{center}
\end{figure}

\par
Also consistent through all simulations is the dominance of evaporation at low- and mid-latitudes. This is attributable to the surface insolation being strongly peaked at the equator in the annual mean, under all considered parameters. Titan's effective obliquity, which over the last 45 kyr has oscillated between 26.4$^{\circ}$ and 28.3$^{\circ}$, would have to increase significantly for the situation to be otherwise. Thus, the net effect is methane accumulation at the poles at the expense of the low- and mid-latitude surface reservoirs.
\par
The more interesting question is what happens to the polar reservoirs with respect to each other.  Figure \ref{Fig:sum_surf_liq} shows the trends in the total surface methane reservoirs divided into five latitude bins, for the duration of the simulations. In all cases, it is immediately apparent that the polar reservoirs initially gain methane as other latitudes lose it, as described above. After about ten Titan years, differences appear that are directly attributable to the different insolations.
\par
The present-day northern polar reservoir continues accumulating methane throughout the simulation. Though the southern polar reservoir at first grows more quickly, it begins to lose methane when the lower latitudes are exhausted. Precipitation is sparse and cannot replenish the loss due to evaporation. In the north, precipitation continues to be sustained and outpaces or balances evaporation. Thus, there is a net flux from south to north pole, and the northern reservoir contains about 50$\%$ more methane by the end of the simulation. 
\par
In direct contrast, the simulation for 28 kyr ago produces the opposite asymmetry, as would be expected given the patterns of insolation. This confirms that indeed the cause of the asymmetries is the forcing, and not some numerical effect. The enhancement of the southern reservoir over the northern one in this case is about 35$\%$, in agreement with the slightly lower polar insolation ratio as compared to the present-day case.
\par
The simulation of 14 kyr ago also produces an enhancement of the northern polar reservoir over the south, though it is less pronounced than in either of the above cases. This is expected, as the insolation difference between the poles in this case is significantly smaller; this suggests that even a small asymmetry can drive a difference in the distribution of liquids, making this highly sensitive to the orbital configuration. Furthermore, it means the north has been gaining surface methane with respect to the south for longer than 14 kyr (this is further discussed in Section~\ref{Sec:discussion}).
\par
The last case provides further confirmation of this mechanism, as both high- and mid-latitude reservoirs evolve equally, under forcing that is practically symmetrical about the equator. Not surprisingly, since the occurrence of surface methane in the model is dominated by polar activity, the asymmetry between low-latitude insolation at opposite equinoxes is irrelevant to the simulated methane distributions.
\par
It is interesting to note that in all cases, once the mid- and low-latitude surfaces have effectively dried, the climate falls into a steadier evolution, and that in all cases this includes episodic precipitation at lower latitudes, seen in Fig.~\ref{Fig:sum_surf_liq} as bumps of similar magnitude in the surface reservoirs of these latitudes. These low-latitude rains occur most  consistently around equinoxes (Supplementary material, Fig.~S5), coinciding with the location of mean meridional upwelling, and consistent with Cassini's observation of equinoctal rainclouds in 2010 \citep{Turtle11b}, as well as the presence of fluvial channels at Huygens' landing site \citep{Tomasko05}. Their occurrence in all four simulations suggests that fluvial erosion, responsible for these channels, is currently active and has been for tens of thousands of years. There is therefore no need to invoke a past, wetter climate as an explanation for the channels near the Huygens landing site.
\par
High-latitude summertime precipitation is persistent throughout the simulations, though the amount varies from year to year, and becomes more sporadic as the surface dries. When the low-latitude surface is nearly dry, it is the polar reservoirs that resupply the atmosphere with methane.

\subsection{Surface temperatures}
\begin{figure}[h!]
	\begin{center}
	\includegraphics[width=0.8\textwidth]{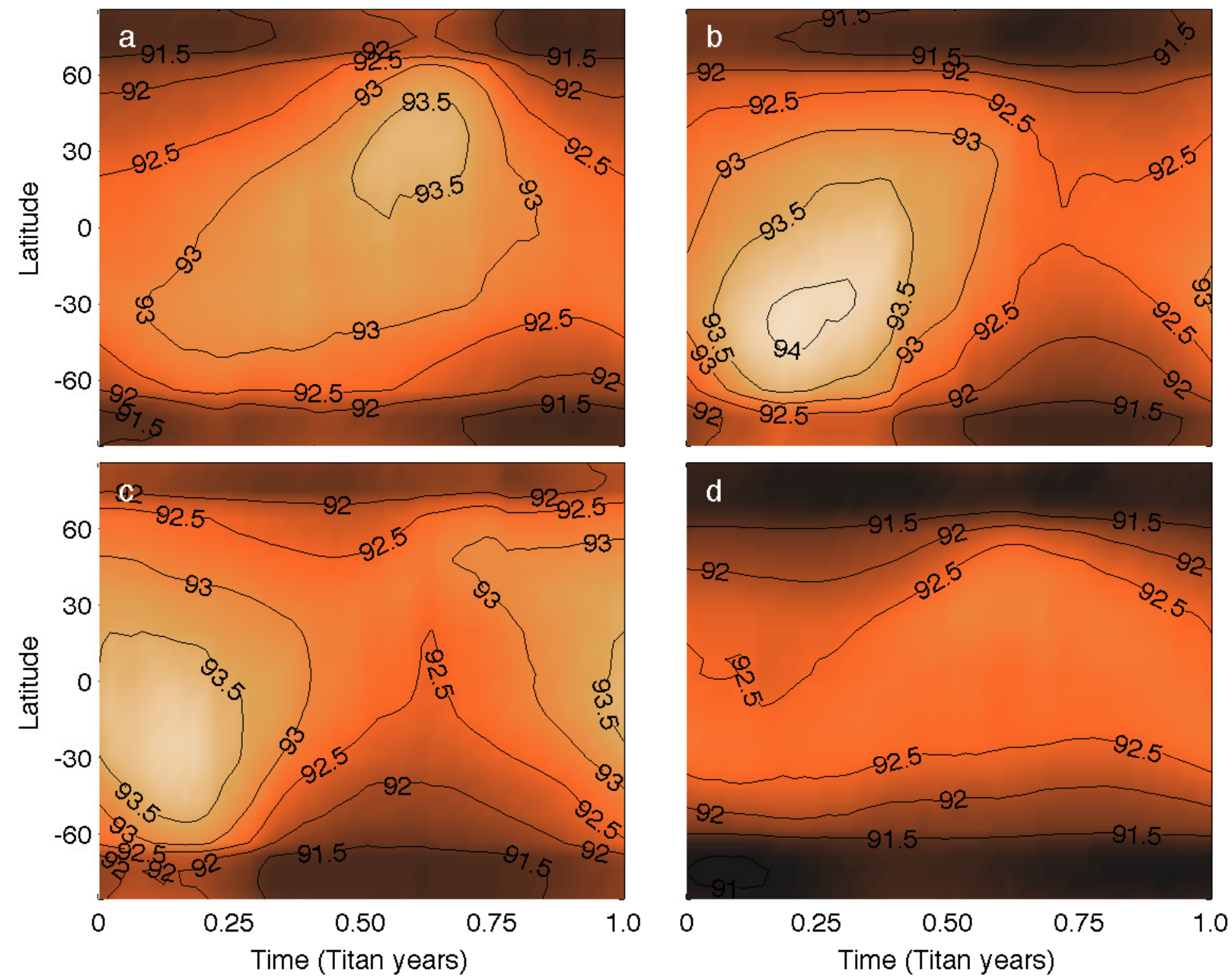}
	\caption[Surface Temperatures]{Surface temperatures (K) for the four simulation cases; (a)-(d) correspond to those in Fig. \ref{Fig:sw_toa}.\label{Fig:tsurf}}
	\end{center}
\end{figure}

Figure \ref{Fig:tsurf} shows the surface temperature variation over a Titan year, averaged for the last ten Titan years of simulation. Interestingly, in both the present-day and 28 kyr ago cases, the temperatures peak at mid-latitudes just before solstice in the hemisphere with the milder summer. However, the other hemisphere remains warmer over the Titan year. At the poles, the surface temperatures are depressed year-round by the presence of liquids. Similarly, the temperature peak coincides with the time and place, in an average sense, of least surface liquid buildup.
\par
The largest gradients in surface temperature appear in the 14 kyr ago simulation, and in this case the peak occurs shortly after perihelion. On the other hand, seasonal changes are muted in the 42 kyr ago simulation, and the overall temperature is lowest in this case. The latter phenomenon is due to the orbital distance being greatest at this time. The former is a consequence of the differences in eccentricity; the high eccentricity 14 kyr ago results in strong differences in insolation between hemispheres that disappear with the very low eccentricity of 42 kyr ago.

\subsection{Circulation and energy transport}
Unsurprisingly, given that the atmospheric structure is unchanged between simulations, the average vertical temperature and wind profiles are largely indistinguishable between the four cases (Fig.~\ref{Fig:temp_wind}). In all cases, a global Hadley-type cell dominates the meridional circulation, with upward motion in the low latitudes and widespread descent elsewhere during equinoxes, and upwelling over summer mid- and high-latitudes and descent in the winter hemisphere during solstices---with the additional development of the summertime polar cell discussed above (Supplementary material, Fig.~S1--S4). The timing of these depends on the seasonal distribution of the insolation, and therefore does vary between simulations.
\par
During the last ten Titan years of simulation, the zonal winds of the lowest model layer, at about 1440 mbar, are weakly westward ($\sim$10 cm/s) at latitudes below $\sim$30$^{\circ}$ in all simulations, except sporadically during equinoxes, possibly accounting for the presence of longitudinal dunes \citep{Tokano10}. Poleward of $\sim$60$^{\circ}$, these winds are prograde year-round, with magnitudes of slightly less than 1 m/s at most times except the $\sim$90$^{\circ}$ of $L_S$ bracketing summer solstice, when they are about twice as strong. In the present-day and 28 kyr ago cases, this enhancement of summertime polar winds is significantly weaker in the pole with the more intense summer (south and north, respectively), while in the 14 kyr ago simulation the north sustains stronger winds throughout the year. The midlatitudes in all cases oscillate between weak eastward and westward winds during spring/summer and fall/winter. 

\begin{figure}[h!]
	\begin{center}
	\includegraphics[width=0.8\textwidth]{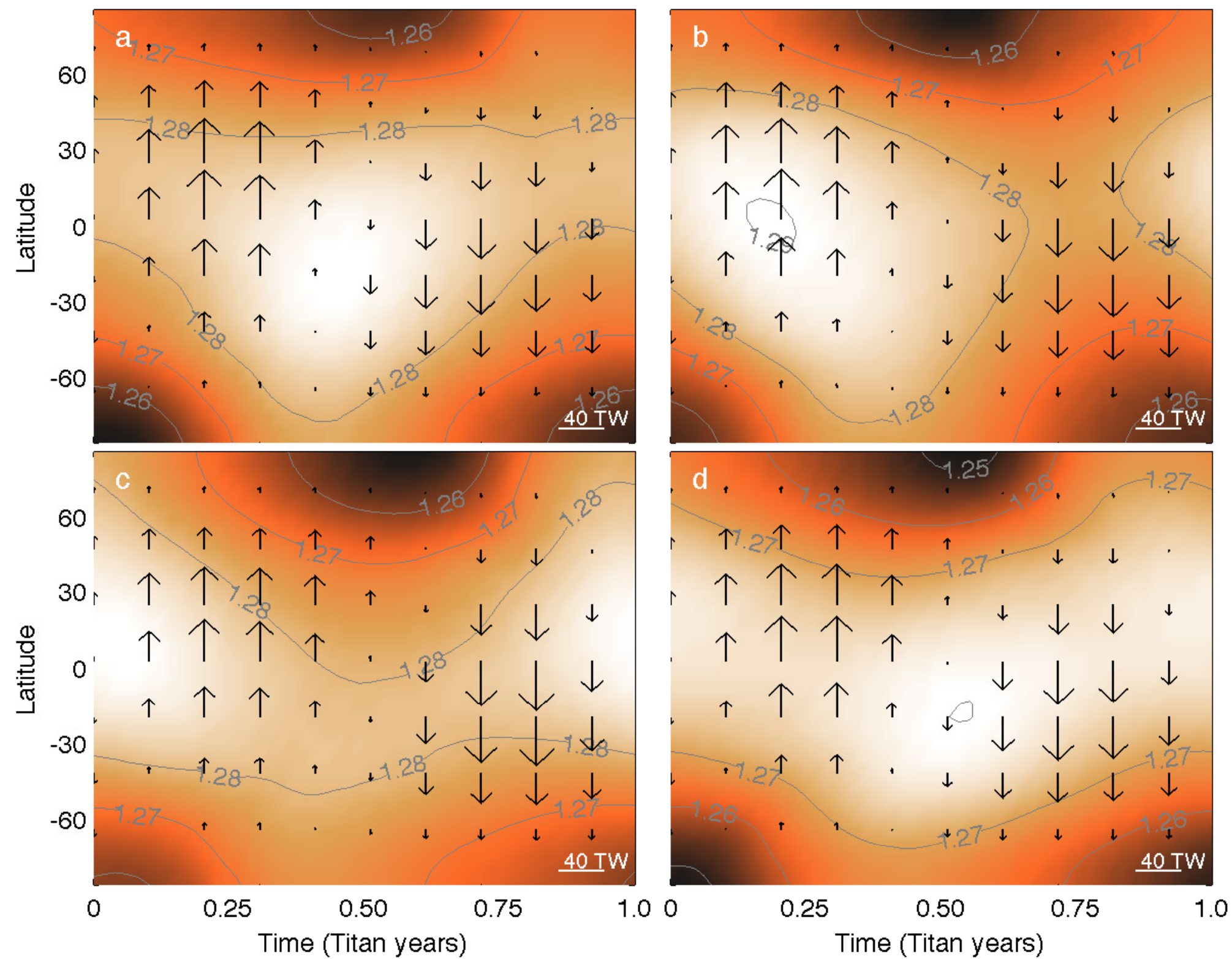}
	\caption[MSE]{Moist static energy (MSE; colors and contours) and MSE fluxes (arrows) for the four simulated cases; (a)-(d) correspond to those in Fig. \ref{Fig:sw_toa}. The units of the MSE are 10$^{10}$J m$^{-2}$; the MSE fluxes are in units of TW. The white lines, with a magnitude of 40 TW, are shown for reference.\label{Fig:MSE}}
	\end{center}
\end{figure}

\par
We can examine the distribution of moist static energy (MSE) as a diagnostic of the (meridional) energy transport by the atmosphere, and its dependence on the insolation. Figure \ref{Fig:MSE} shows the MSE in colors, with gray contours, for results averaged over the last ten Titan years of simulations for the four cases. The overlaid arrows indicate the magnitude and direction of the flux of MSE. An immediately apparent feature of these results is that, in contrast to the model of \citet{Schneider12}, the MSE maximum never reaches the polar regions in any simulation. This is directly attributable to Schneider et al's use of a simplified radiative transfer scheme that incorrectly calculates the insolation distribution in the troposphere and at the surface \citep[see][]{Lora11}. A consequence of this is that polar precipitation begins in late spring (Section~\ref{Sec:precip}).
\par
In both the present-day and 28 kyr ago cases, the MSE peaks shortly after solstice in the hemisphere with the more intense summer. The MSE flux is directed into the winter hemisphere. As in \citet{Mitchell12}, we find that the flux of latent energy is directed $opposite$ to the MSE flux (Supplementary material, Fig.~S6); thus, methane is advected for a longer time into the hemisphere with the longer, weaker summer.  For the 14 kyr ago and 42 kyr ago simulations, the MSE maximum also responds to the timing of perihelion, occurring shortly after it. Because the MSE is more latitudinally symmetric, the fluxes into and out of both hemispheres are also more balanced.

\begin{figure}
	\begin{center}
	\includegraphics[width=0.6\textwidth]{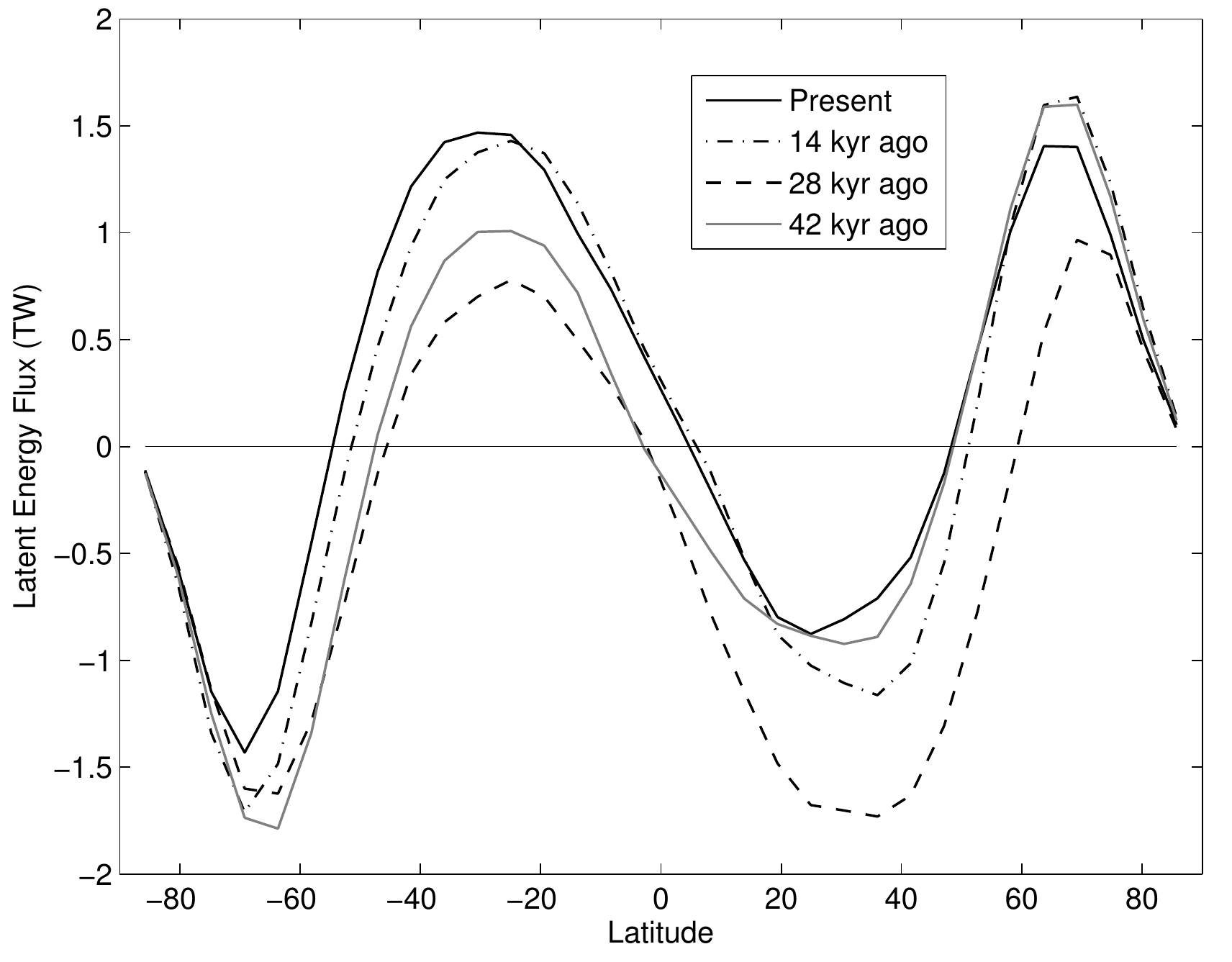}
	\caption[LE Fluxes]{Annual-mean latent energy fluxes for the four simulations.\label{Fig:LEflux}}
	\end{center}
\end{figure}

\par
Figure \ref{Fig:LEflux} shows the annual-mean latent energy fluxes, averaged over the last ten Titan years of simulation. In all cases, the atmosphere converges methane at or close to the equator. This is exactly opposite to the results of \citet{Mitchell09} and \citet{Mitchell12}, perhaps also as a consequence of the details of the radiative transfer. This convergence---though it is a feature only of the annual-mean fluxes, as instantaneous latent energy fluxes are $across$ the equator---may provide the methane necessary for the episodic equatorial precipitation discussed above. The annual-mean dry static energy (DSE) flux in all cases is divergent at the equator and peaks at mid-latitudes with a magnitude of $\sim$10 TW (Supplementary material, Fig.~S7). The latent energy flux is therefore only dominant in the MSE at higher latitudes, where surface moisture is perpetually available.
\par
The differences due to the contrasting patterns of insolation can be seen most clearly in the latent energy fluxes at the mid-latitudes. The imbalance in this transport between hemispheres is indicative of where surface methane preferentially builds: Northward transport in the south is larger than southward transport in the north in the present-day case, and the opposite is true of the 28 kyr ago simulation; the case 42 kyr ago is nearly symmetrical around the equator.

\begin{figure}[h!]
	\begin{center}
	\includegraphics[width=0.6\textwidth]{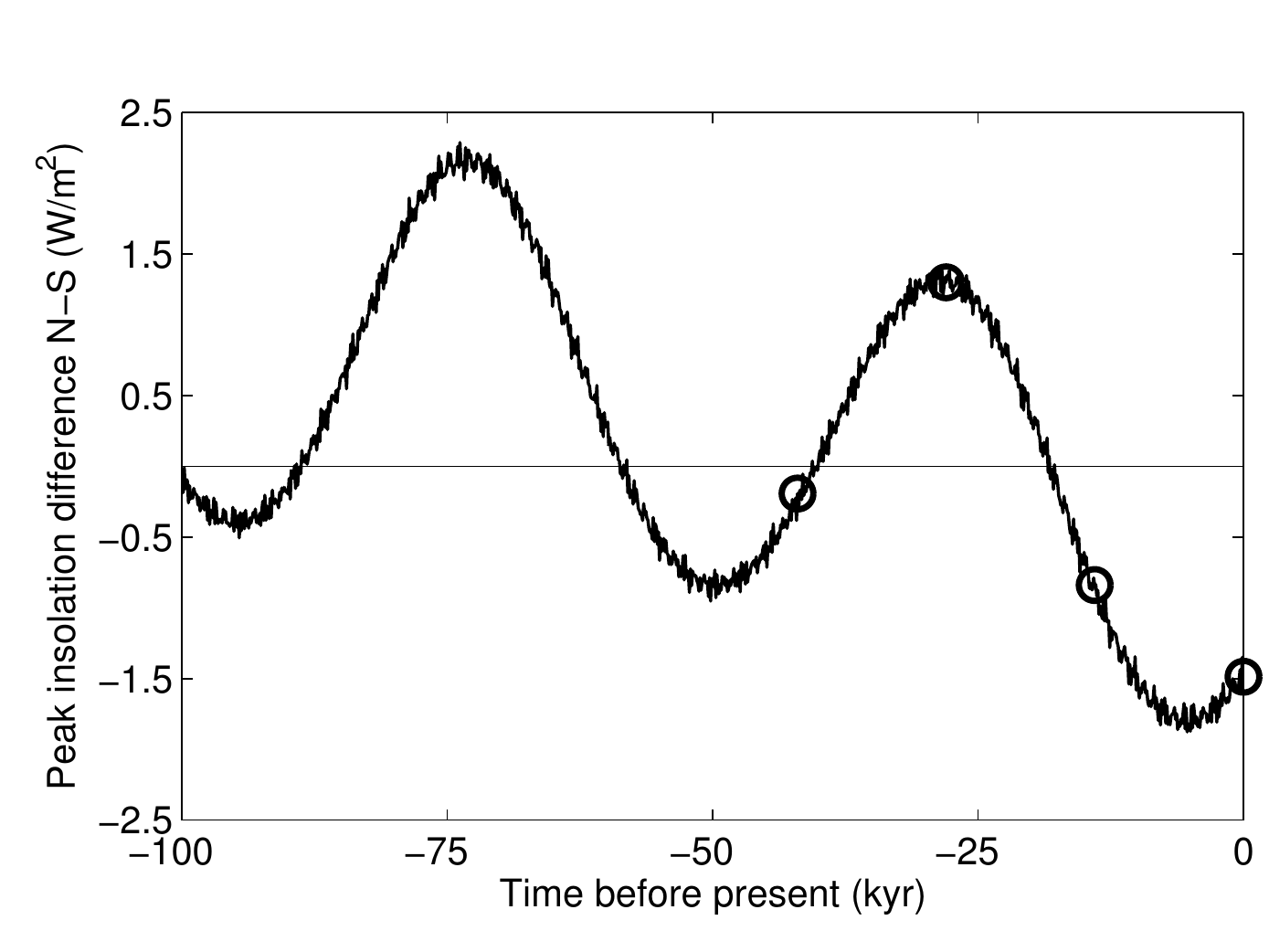}
	\caption[Insolation difference]{Difference in the peak annual insolation between north and south for the past 100 kyr. The times and corresponding values of the GCM simulations are marked by circles.}\label{Fig:insol_diff}
	\end{center}
\end{figure}

\subsection{Transport over long timescales}
\begin{figure}[h!]
	\begin{center}
	\includegraphics[width=0.6\textwidth]{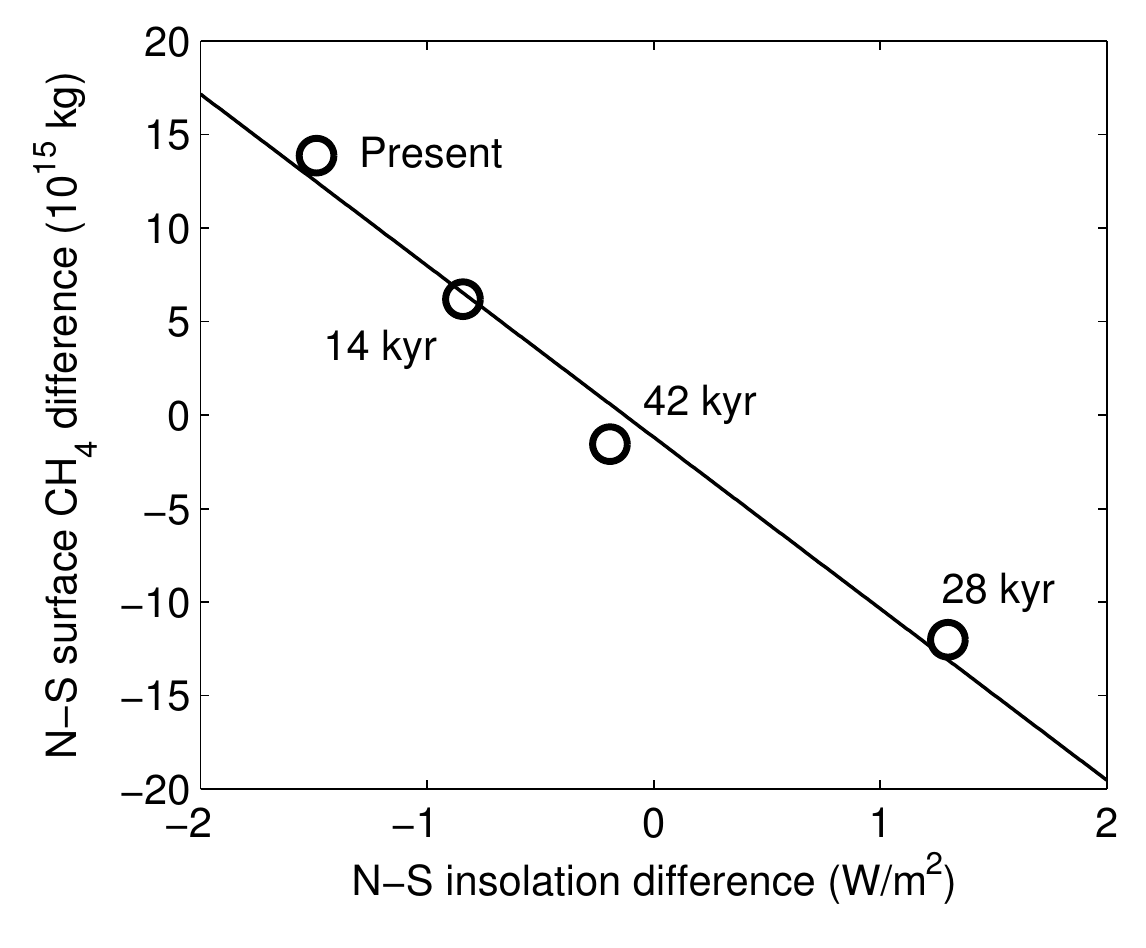}
	\caption[N-S differences]{Relationship between the peak N-S insolation differences (Fig.~\ref{Fig:insol_diff}) and the net N-S differences in polar reservoirs at the end of the simulations. The circles mark GCM values, which follow on a linear trend.}\label{Fig:ch4_diffs}
	\end{center}
\end{figure}

\citet{Aharonson09} suggested that the top-of-atmosphere radiation asymmetry results in the asymmetry of evaporation minus precipitation, such that the north-south difference in peak annual insolation might be directly relatable to the transport of methane. Figure~\ref{Fig:insol_diff} shows this difference for the polar regions over the past 100 kyr. We compare the north-south difference in peak insolation and the north-south difference in the content of the polar surface reservoirs for the four simulations in Fig.~\ref{Fig:ch4_diffs}. This indicates a significant correlation between the two, showing that the orbital forcing does impact the surface liquid distribution, in agreement with the hypothesis of \citet{Aharonson09}.

\par
\begin{figure}[h!]
	\begin{center}
	\includegraphics[width=0.6\textwidth]{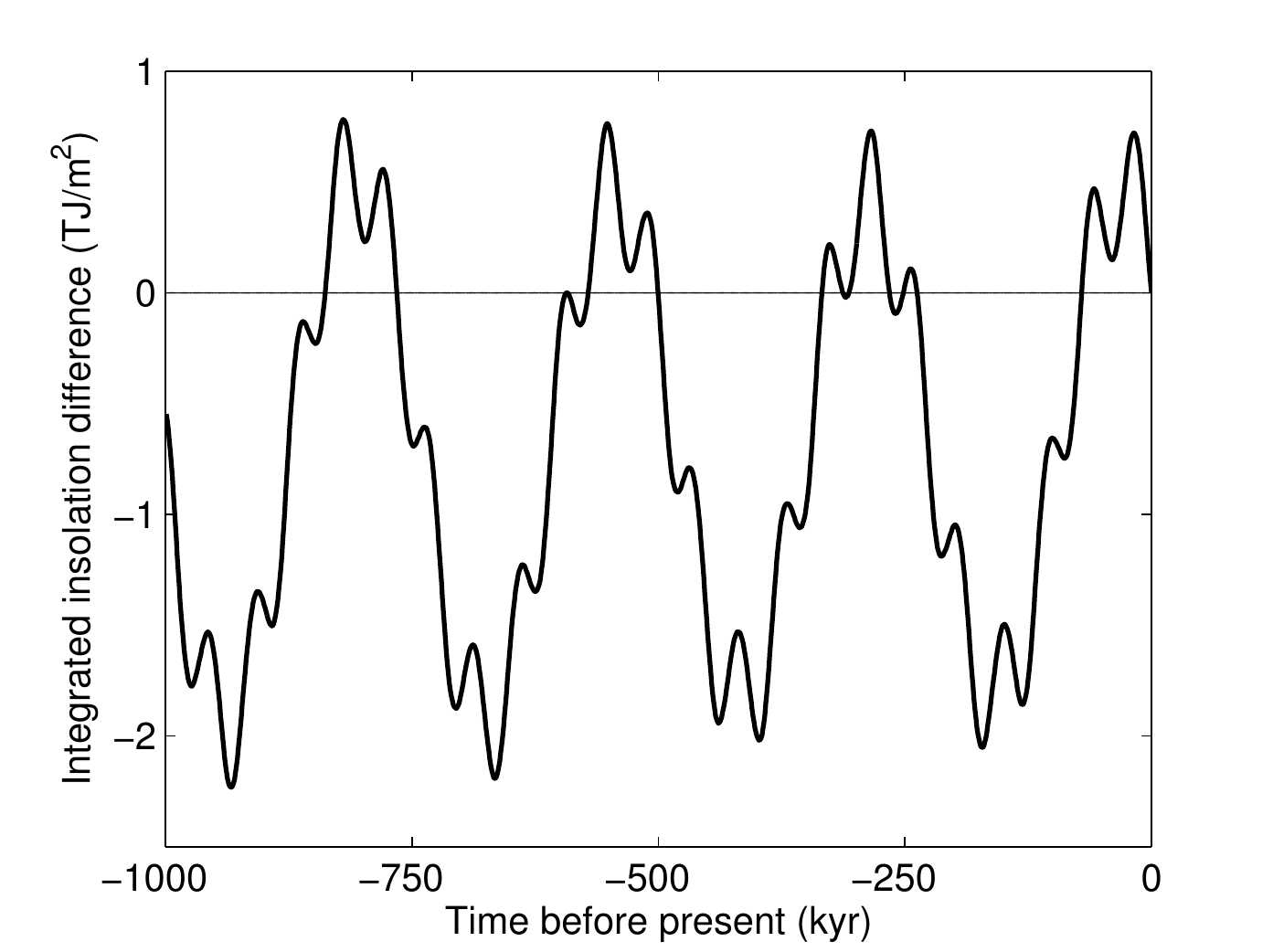}
	\caption[Integrated insolation]{Relative net integrated insolation difference going back one million years. Note that values are relative to the present day, and positive values correspond to northward \textit{transport}; i.e., negative N-S insolation differences. The long-term oscillation occurs around a constant value, indicating no long-term net transport.}\label{Fig:int_insol}
	\end{center}
\end{figure}

Having established that the orbital forcing can induce an asymmetry in the surface liquid distribution, a basic estimate of the timescale for transporting the surface reservoir from pole to pole can be made based simply on the insolation asymmetry. The peak insolation differences over the past million years are calculated from the top-of-atmosphere insolation distributions (see Appendix), and these can then be integrated backward in time (Fig.~\ref{Fig:int_insol}). The result is a curve of net energy---a proxy for pole-to-pole transport \citep{Aharonson09}---with a dominant period of about 125 kyr, and no net or long-term trend (the long-period oscillations are about a constant value). This implies that asymmetric transport due to the orbital forcing can only be manifested as an asymmetrical surface liquid distribution if the timescale to transport the surface liquid reservoir is less than about 125 kyr.

\section{Discussion and implications}\label{Sec:discussion}
The simulation results show that methane is, on average, transported poleward under all conditions considered. This is not in itself surprising, but these conditions include a wide range of eccentricity and longitude of perihelion variations. Apsidal precession is the dominant mechanism controlling long-term changes in forcing: Titan's effective obliquity, which controls the annual-mean latitudinal distribution of insolation, varies much less. Indeed, over the past million years, it has ranged between 26.3$^{\circ}$ approximately 310 kyr ago, and 28.4$^{\circ}$ 640 kyr ago. Therefore, Titan's polar regions have likely been gaining or retaining liquids, at the expense of the lower latitudes, for at least the last million years. Though occasional precipitation forms at most latitudes, the low- and mid-latitudes have likely been desert-like for as long.
\par
The results also confirm that orbital forcing can induce an asymmetry in the surface liquid distribution. Furthermore, the simulations suggest that 14 kyr ago, methane was transported from south to north on average, as it is in the present day, though with a lower magnitude; around 28 kyr ago, the transport was reversed, but 42 kyr ago there was very little asymmetry. Thus, the net poleward transport of surface methane over the past 42 kyr has only been slightly northward. This is confirmed in Fig.~\ref{Fig:int_insol} (assuming that peak insolation differences are indeed a good proxy for atmospheric methane transport), which further suggests that the timescale for transporting the total surface reservoir of liquid methane is relatively short, given the large currently-observed dichotomy.
\par
In order to properly reconcile the insolation differences over the last 100kyr and the implied transport (Figs.~\ref{Fig:insol_diff} and \ref{Fig:int_insol}) with the observed large present-day dichotomy, the relevant timescale for transporting the surface methane reservoir should be $\lesssim$30 kyr, which corresponds to a large positive value of the integrated insolation difference, relative to present, at 30 kyr ago in Fig.~\ref{Fig:int_insol}. If this is the the case, then the asymmetry reverses completely with a period of $\sim$125 kyr. Otherwise, there is either insufficient time to generate the large observed asymmetry (and the orbital forcing is only a secondary component of generating the asymmetry), or, for timescales $\gtrsim$125 kyr, the orbital forcing produces no net asymmetry, in which case the observed dichotomy would be ancient and due primarily to a different mechanism.
\par
It is important to note that uptake of the surface reservoir into the atmosphere could be throttled by a variety of processes not included in the simulations. Some examples would be a significant fraction of ethane in the liquid, deep reservoirs with small surface areas, or significant and efficient infiltration of precipitation into the sub-surface, all of which would act to lower the mean rate of evaporation and reduce the reservoir-atmosphere interaction. Furthermore, several physical processes in the model are treated in an idealized fashion, limiting the accuracy of the finer-scale results. It is therefore unreasonable to attempt to directly quantify the transport timescales discussed above using GCM output, especially given the snap-shot nature of the simulations. Further model development is necessary to investigate this more quantitatively.
\par
Finally, because the cold trap of the tropopause is not very effective (unlike on Earth), especially at low latitudes, methane leaks into the stratosphere, where it is effectively removed from the tropospheric methane cycle. If some methane is constantly evaporating, this implies a net loss from the surface to the atmosphere. On Titan, this methane is destroyed by photolysis in the upper atmosphere; the puzzle of its replenishment is not a new one. 
\par
It is worth noting that during the moist spin-up phase in our modeling there is too much methane at the surface for the low- and mid-latitudes to dry out. However, methane is still transported poleward by the atmosphere, with a net gain in the surface reservoirs poleward of 60$^{\circ}$. Thus, the model results reported here are independent of the actual magnitude of the total surface-atmosphere reservoir of methane, as long as there is enough to sufficiently humidify the atmosphere. A much larger reservoir would result in the same relative distributions of surface liquids, but over a longer timespan that makes modeling it prohibitively time consuming (hence our approach).

\section{Summary}
Using a new general circulation model for Titan that includes a detailed treatment of radiative transfer, we have simulated the climate of Titan under four characteristic orbital configurations that occurred over the past 42 kyr. The results indicate that Titan's atmosphere, barring significant changes in composition not considered here, is to first-order insensitive to changes in Saturn's orbit. In general, the atmosphere efficiently transports methane poleward, so that surface liquid buildup in polar regions is persistent and lower latitudes are enduring deserts with occasional rainfall.
\par
In detail, however, seasonal asymmetries do affect the distribution of methane, and surface liquids migrate to the pole that experiences less intense summers. For at least the past 14 kyr, the net transport of methane has been northward, potentially explaining the great dichotomy of observed lakes on Titan if the timescale to move the entire reservoir is of that order. On the other hand, if the timescale is long, the observed dichotomy may be ancient and due primarily to another mechanism.

\section*{Acknowledgements} 
J.M.L. gratefully acknowledges support from NASA Earth and Space Science Fellowship NNX12AN79H, as well as the Cassini Project. J.I.L. was supported by the Cassini Project. Simulations were carried out with an allocation of computing time on the High Performance Computing systems at the University of Arizona. The authors thank two anonymous reviewers whose comments greatly improved and clarified the manuscript. 

\section*{Appendix A}
A 1.0 Myr integration of the SWIFT orbital position integration code \citep[\url{http://www.boulder.swri.edu/~hal/swift.html};][]{Levison94} was used to generate Saturn's inclination $i$, longitude of ascending node $\Omega$, argument of periapsis $\omega$, longitude of perihelion $\varpi$, and eccentricity $e$ over this time. 
\par
Obliquity variations were calculated using secular theory, accounting for the spin-orbit resonance with Neptune \citep{Ward04}: The secular obliquity variation can be written, accurate to first order in inclination, as \citep[Eq. 32 in][]{Ward74}
\begin{equation}
\theta \approx \bar{\theta} - \sum_j\frac{g_jI_j}{\alpha\cos{\bar{\theta}}+g_j}\sin{(\alpha t \cos{\bar{\theta}} + g_jt + \Delta_j - \Phi - \Omega_0)} - \sin{(\Delta_j- \Phi - \Omega_0)},
\end{equation}
where $\bar{\theta}$ is a long-term average obliquity (26.73$^{\circ}$, J2000), $g_j$ and $I_j$ are the three largest-amplitude terms for Saturn's orbit, $t$ is time, $\Delta_i$ are phase constants, and $\Phi=-30.06^{\circ}$ and $\Omega_0=113.67^{\circ}$ are the initial pole azimuth and longitude of the ascending node, respectively. Values for $g_j$ and $I_j$ are given in Table 1 of \citet{Ward04} and $\Delta_j$ are taken from \citet{Bretagnon74}. The precessional constant $\alpha$ is \citep[Eq. 1 in][]{Ward04}:
\begin{equation}
\alpha = \frac{3}{2}\frac{n^2}{f}\frac{J_2+q}{\lambda+l},
\end{equation}
where $f$ is the spin frequency of Saturn, $n$ its heliocentric mean motion, $J_2$ the coefficient of the quadrupole moment of its gravity field, $\lambda$ its normalized moment of inertia, $q$ the effective quadrupole coefficient of the satellite system, and $l$ the normalized angular momentum of the satellite system. Following \citet{Ward04}, $\alpha=0.''8306$ yr$^{-1}$.
\par
The variation in azimuth of the spin pole $\varphi$ is given by \citep[Eq. 36 in][]{Ward74}
\begin{equation}
\varphi = \Phi - \alpha t \cos{\bar{\theta}^{\star}} - \Omega + \Omega_0 + \sum_j I_j^{'}[\cos{(\alpha t \cos{\bar{\theta}} + g_jt + \Delta_j - \Phi - \Omega_0)} - \cos{(\Delta_j - \Phi - \Omega_0)}],
\end{equation}
where
\begin{equation}
I_j^{'} = \frac{g_jI_j}{(\alpha\cos{\bar{\theta}}+g_j)\sin{\bar{\theta}}}\left(\frac{g_j\cos{\bar{\theta}}+\alpha}{g_j+\alpha\cos{\bar{\theta}}}\right),
\end{equation}
and
\begin{equation}
\bar{\theta}^{\star} = \bar{\theta} + \sum_j\frac{g_jI_j}{\alpha\cos{\bar{\theta}}+g_j}\sin{(\Delta_j - \Phi - \Omega_0)}.
\end{equation}
\par
Finally, the solar longitude of perihelion $L_{S,p}$ is given by $L_{S,p}=\omega - \varphi - 90^{\circ}$.
The top-of-atmosphere diurnal-mean insolation distribution for any epoch can be calculated given $e$, $\theta$, and $L_{S,p}$ \citep{Berger78}.

\end{document}